\documentclass[%
reprint,
nofootinbib,
superscriptaddress,
aps,
prx,
floatfix,
]{revtex4-2}

\usepackage{braket}
\usepackage{amsmath}
\usepackage{amsthm}
\usepackage{amssymb}
\usepackage{amsfonts}
\usepackage{mathtools} 
\usepackage{graphicx}
\graphicspath{{images/}}
\usepackage{dcolumn}
\usepackage{bm}

\usepackage{float} 
\usepackage{lipsum} 

\usepackage{mathpazo} 
\usepackage{avant} 

\usepackage[para]{threeparttable}
\usepackage{array,booktabs,longtable,tabularx,tabulary}
\newcolumntype{L}{>{\raggedright\arraybackslash}X}
\usepackage{ltablex}

\usepackage{siunitx,booktabs} 
\AtBeginDocument{
\heavyrulewidth=.08em
\lightrulewidth=.05em
\cmidrulewidth=.03em
\belowrulesep=.65ex
\belowbottomsep=0pt
\aboverulesep=.4ex
\abovetopsep=0pt
\cmidrulesep=\doublerulesep
\cmidrulekern=.5em
\defaultaddspace=.5em
}

\usepackage[flushleft]{threeparttablex}
\usepackage{array}
\newcolumntype{P}[1]{>{\centering\arraybackslash}p{#1}}

\usepackage{chngcntr}

\usepackage{enumerate}

\usepackage{hyperref}
\usepackage{cleveref} 

\hypersetup{
    colorlinks=true,
    linkcolor=blue,
    filecolor=blue,
    urlcolor=blue,
    citecolor=blue,
}
\usepackage[dvipsnames]{xcolor}

\usepackage{appendix}
\usepackage{amsmath,amsfonts}
\allowdisplaybreaks

\usepackage{xfrac}

\usepackage{tikz} 

\usepackage{centernot}

\usepackage[most]{tcolorbox} 

\definecolor{redd}{RGB}{208, 2, 27}
\definecolor{bluu}{RGB}{74, 144, 226}
\definecolor{my-yellow}{RGB}{245,166,35}
\definecolor{yelloww}{RGB}{248,231,28}

\begin{document}

\title{Information flow-enhanced precision in collisional quantum thermometry   }

\author{Taysa M. Mendon\c{c}a}
\email{tmendonca@ifsc.usp.br}
\affiliation{Instituto de F\'{i}sica de S\~{a}o Carlos,
Universidade de S\~{a}o Paulo, CP 369, 13560-970, S\~{a}o Carlos, SP, Brazil}
\affiliation{Centre for Quantum Materials and Technologies, School of Mathematics and Physics, Queen’s University, Belfast BT7 1NN, United Kingdom}

\author{Diogo O. Soares-Pinto}
\email{dosp@ifsc.usp.br}
\affiliation{Instituto de F\'{i}sica de S\~{a}o Carlos,
Universidade de S\~{a}o Paulo, CP 369, 13560-970, S\~{a}o Carlos, SP, Brazil}

\author{Mauro Paternostro}
\email{mauro.paternostro@unipa.it}
\affiliation{Universit\`a degli Studi di Palermo, Dipartimento di Fisica e Chimica - Emilio Segr\`e, via Archirafi 36, I-90123 Palermo, Italy}
\affiliation{Centre for Quantum Materials and Technologies, School of Mathematics and Physics, Queen’s University, Belfast BT7 1NN, United Kingdom}

\begin{abstract}

We describe and analyze a quantum thermometer based on a multi-layered collisional model. The proposed architecture provides significant sensitivity even for short interaction times between the ancillae comprised in the thermometer and the system to be probed, and a small number of information-acquiring collisions. The assessment of the flow of information taking place within the layered thermometer and between system and thermometer reveals that the tuning of the mutual backflow of information has a positive influence on the precision of thermometry, and helps unveiling the information-theoretic mechanisms behind the working principles of the proposed architecture.

\end{abstract}
 
\pacs{Pacs}

\maketitle

\section{Introduction}

Quantum metrology uses resources such as entanglement \cite{Giovannetti2006, Maccone2013,Maccone2016}, quantum coherence~\cite{Serra2015a,Pires2018,Castellini2019} and squeezing~\cite{Maccone2020} to enhance the precision with which measurement protocols are able to estimate relevant physical quantities beyond the capabilities of classical strategies. When applied to the problem of estimating the temperature of a system, such quantum tools make {\it quantum thermometry}~\cite{DePasquale2018} effective even when operating under conditions that are classically challenging, such as when the thermometer has not yet thermalized with the system. 
This makes such framework potentially useful to probe non-equilibrium environments, which could showcase coherence and correlations. 

Energetic resource value of information is an active research area in quantum thermodynamics, they can power up, cool, or heat systems using information instead of consuming energy. Quantum thermometry will be an integral part of practical implementations of such quantum information engines and devices.
A particularly powerful way to model the process of estimating temperature leverages upon the framework of \textit{collisional models}~\cite{CICCARELLO2021,Ciccarello2017}, which has been successfully applied to describe the thermodynamics of non-equilibrium quantum processes~\cite{Paternostro2019b, Colisao_Filippov2022, ColisaoBateria_Landi2021, Paternostro2014c, ColisaoEstoc_Steve2021, Lutz2020}.

In collisional quantum thermometry, a set of probing particles acquire information on the temperature of the system of interest through a sequence of short and intense, crash-like interactions. Ref.~\cite{Landi2019_Colisao} explored such approach showing how it allows for the estimation of the temperature of a quantum system beyond the precision constraints set by thermal fluctuations~\cite{Colisao_Paris}. Additional studies have grounded the suitability of collisional quantum thermometry to pursue this important metrological task~\cite{ColisaoEstoc_Steve2021}. 
In this paper, we use a collision-based approach to build a thermometric sensor that measures the temperature of an environment indirectly, without thermalization between sensor and system, using measurements performed on auxiliary qubits (sensor) organized in the form of layers of chains. We go beyond state-of-the-art in collisional quantum thermometry by allowing explicitly for inter-layer interactions: although this obviously increases the complexity of the problem at hand, it also enhances the process of information-spreading that is responsible for the performance of this metrological task. Quantitatively, we achieve a thermometric sensitivity  that is orders of magnitude larger than that of a single-layered thermometer as proposed by \cite{Landi2019_Colisao}. We also investigate the case where there is a complete exchange of information during the system-ancilla interaction, a model proposed by \cite{Colisao_Scarani}. In this case, we showed that the same amount of information flows between the qubits, not allowing for a gain in sensitivity. 

To investigate the motivating factors for such gain in sensitivity of our multi-layered thermometer, we studied the behavior of the information flow using the measure of non-Markovianity  proposed in Ref.~\cite{Breuer2009,Breuer2010} (dubbed here as BLP), this procedure was demonstrated in Ref.\cite{mendonca2024InfFlow}. Finally, we compare the QFI and the BLP measurement, qubit by qubit, to investigate the path taken by information in our complete system. It was possible to determine how the accumulation of information provides an increase in the sensitivity of our thermometer, we also show that such accumulation can also be characterized through Mutual Information.
\noindent
\section{Description of the Model}
\begin{figure}[t!]
\centering
\includegraphics[width=1.0\columnwidth]{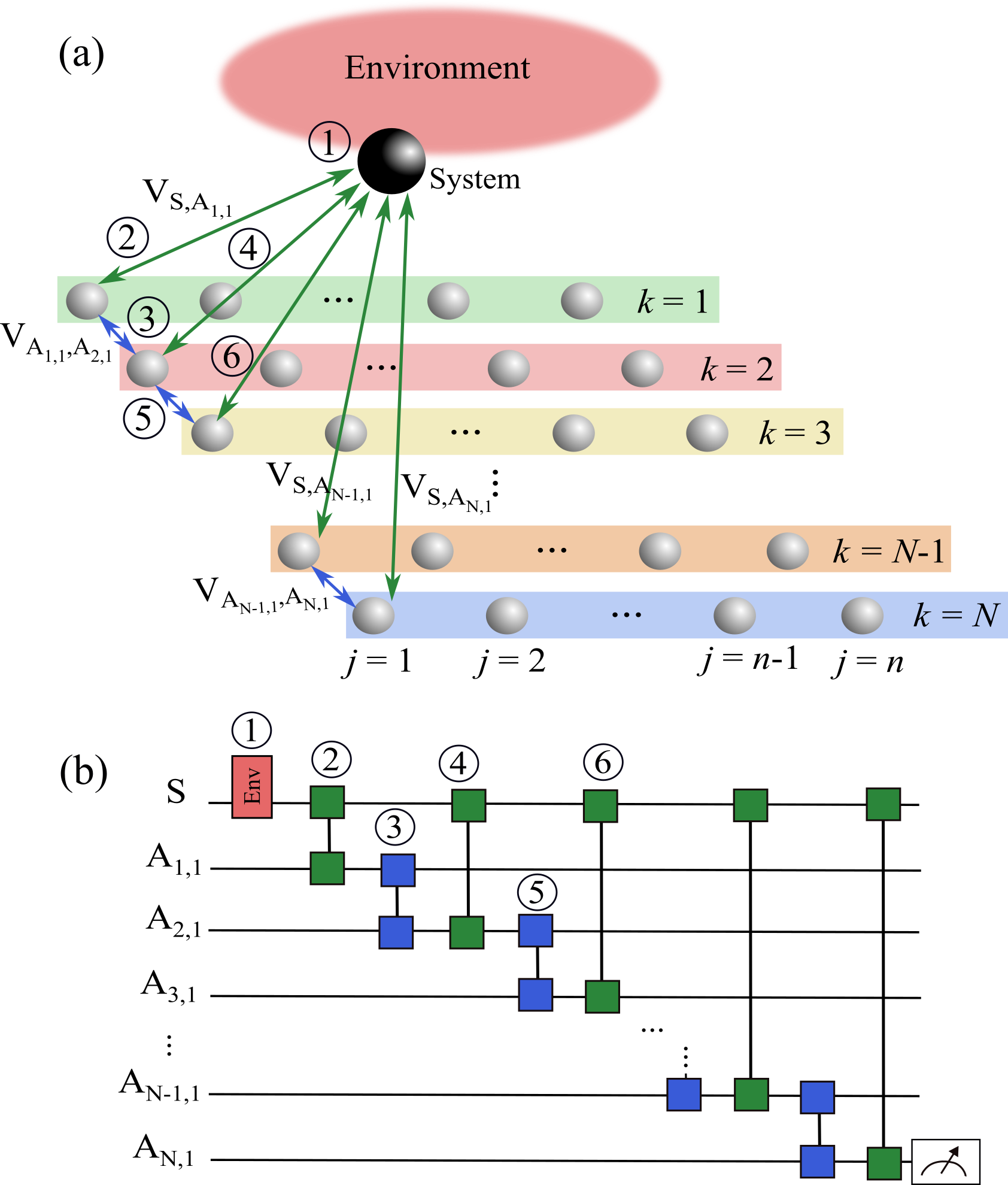}
\caption{(a) Illustration of the scheme of interctions for  multi-layered collisional thermometry. System $S$ thermalizes with an environment in equilibrium at temperature $T$. The inference of such parameter is made through the sequence of {\it collisional} interactions illustrated in the main text. The numbered steps of the scheme [from (1) to (6) in panel (a)] illustrate the first few steps of the sequence leading to the measurements entailed by our metrological approach to the estimate of $T$. 
(b) Circuit-model representation of interactions of the steps (1)-to-(6) of our protocol. Each inter-ancilla collision is depicted by a full phased-SWAP gate [colored in blue], while the green-colored gates stand for  is used for partial SWAPs.}
\label{fig:Sistema}
\end{figure}

We consider a collisional model in which a qubit -- embodying the system -- is initially thermalized with an environment at temperature $T$. A set of ancillary chains is used to indirectly estimate the temperature of the environment, as shown in Fig.~\ref{fig:Sistema}(a). The dynamics of the overall compound comprises a sequence of collisions between the system and $N$ chains of $n$ ancillae each. In a schematic manner, the process unfolds according to the following steps:\\
\noindent
In the first step, the system $S$ thermalizes with the environment through the thermal map $\mathcal{T}_{S}=\mathrm{e}^{\mathcal{L}\tau_{SE}}$  generated by~\cite{PetruccioneLivro2002}
\begin{equation}
	\frac{d\rho_{S}}{dt} = \mathcal{L}(\rho_{S})= \gamma\left(\bar{n}+1\right)\mathcal{D}\left[\mathcal{S}_{-}\right] + \gamma\left(\bar{n}\right)\mathcal{D}\left[\mathcal{S}_{+}\right],
	\label{eq:SE_Termaliza}
\end{equation}
where $\mathcal{D}\left[L\right]=L\rho_{S}L^{\dag}-\frac{1}{2}\left\{L^{\dag}L,\rho_{S}\right\}$, $\gamma$ is a temperature-independent coupling constant, $\bar{n}=(\mathrm{e}^{\hbar\beta\Omega_S}-1)^{-1}$ with $\beta=1/k_BT$, $k_B$ the Boltzmann constant and $T$ is the temperature of the environment. Here,  $\Omega_{S}=1$ is the characteristic frequency of the system and $\tau_{SE}$ is a time long enough for the system qubit to be thermalized with the temperature of the environment. At the end of this step, the qubit system will be in the Gibbs  state $\rho_{S}^{th}=\mathrm{e}^{-H_{S}/k_{B}T}/Z$, where $H_{S}= \hbar\Omega_{S}\mathcal{S}_{z}/2$ is the Hamiltonian of the system.

The following steps involving $A_{k,j}$, that is, the element $j^\text{th}$ of the chain $k^\text{th}$, with $j=1,..,n$ and $k=1,..,N$. 
A collision between the system $S$ and $A_{{k,j}}$, ruled by the Hamiltonian 
\begin{equation}
	\label{eq:HSA}
	H_{SA_{k,j}} = \hbar g \left(\mathcal{S}_{+}\alpha^{k,j}_{-}+ \mathcal{S}_{-}\alpha^{k,j}_{+}\right)
\end{equation}
occurs, followed by the collision between $A_{k,j}$ and $A_{k+1,j}$ according to the generator 
\begin{equation}
	\label{eq:HAA}
	H_{A_{k,j}A_{k+1,j}} = \hbar J \left(\alpha^{k,j}_{+}\alpha^{k+1,j}_{-}+ \alpha^{k,j}_{-}\alpha^{k+1,j}_{+}\right).
\end{equation}
Here, $g=1$ ($J=1$) is the system-ancilla (inter-ancilla) coupling strength, which we assume to be independent of $j$ and $k$, while the operators $\mathcal{S}_{+}=(|{1}\rangle\langle{0}|)_S$, $\mathcal{S}_{-}=(|{0}\rangle\langle{1}|)_S$ and $\alpha_{+}^{k,j}=(|{1}\rangle\langle{0}|)_{A_{k,j}}$, $\alpha_{+}^{k,j}=(|{0}\rangle\langle{1}|)_{A_{k,j}}$ refer to system and  the relevant ancilla, respectively, with $\{|{0}\rangle,|{1}\rangle\}$ being the computational basis of our problem (we have shifted the energy of the states in such a way that $\mathcal{O}_{z}|{0}\rangle=|{0}\rangle$ and $\mathcal{O}_{z}|{1}\rangle=-|{1}\rangle$).

The collisions continue so as to span all the chains (i.e. for $k=1,...,N$), after which a thermalization process occurs again. The sequence of system-ancilla and ancilla-ancilla collisions restarts for the $j+1$ ancillae spanning all the $N$ chains. 
Fig. \ref{fig:Sistema} provides a visual depiction of the first few steps of the process, as well as its circuit representation representing the interactions for the case of few chains only.

The sequence of operations that describes the described interaction protocol for a single repetition and considering a generic ancilla leads to the state
\begin{eqnarray}
	\label{eq.Oper_U}
	\rho_{S,A_{1,j}\cdots A_{N,j}}{=} & \mathcal{U}_{S,A_{N,j}}\circ\mathcal{U}_{A_{N-1,j}A_{N,j}}\circ\dots \nonumber \\  &\circ\mathcal{U}_{S,A_{2,j}}\circ\mathcal{U}_{A_{1,j}A_{2,j}}\circ\mathcal{U}_{S,A_{1,j}}\circ\mathcal{T}_{S}(\rho_0),
\end{eqnarray}
where $\mathcal{U}_{S,A_{k,j}}\circ \mu = V_{S,A_{k,j}}\, \mu\, V_{S,A_{k,j}}^{\dagger}$ and $\mathcal{U}_{A_{k,j},A_{k+1,j}}\circ \mu = V_{A_{k,j},A_{k+1,j}} \,\mu\, V_{A_{k,j},A_{k+1,j}}^{\dagger}$ for a density matrix $\mu$ and  $V_{S,A_{k,j}}=\text{exp}(-i H_{SA_{k,j}}\tau_{SA}/\hbar)$, $V_{A_{k,j},A_{k+1,j}}=\text{exp}(-i H_{A_{k,j}A_{k+1,j}}\tau_{A}/\hbar)$. $\rho_0$ is the initial state such that $\rho_0 = \rho_{S}(0)\otimes \rho_{A_{N,j}}(0)$ where $\rho_{S}(0)=(|{0}\rangle\langle{0}|)_S$ for the system and $\rho_{A_{k,j}}(0)=(|{0}\rangle\langle{0}|)_{A_{k,j}}$ for the ancillae.
In our analysis, we will consider values of the dimensionless system-ancilla  interaction time $g\tau_{SA}\ll{\pi/2}$, here the interaction time between the qubits is equivalent to a non-full swap gate, so the information is not transferred entirely between them, there is no thermalization between the system and the thermometer, there is only a "touch" between them. Now the inter-ancilla is $J\tau_{A}=\pi/2$, this interaction time is equivalent to a full-swap gate, which means that all information is transferred between the qubits. Therefore, under such conditions, a full exchange of information between the ancillae will be allowed, as $V_{A_{k,j}A_{k+1,j}}$ is the phased SWAP gate performing the transformation $|{a,b}\rangle_{A_{k,j}A_{k+1,j}}\to(-i)^{a\oplus b}|{b,a}\rangle_{A_{k,j}A_{k+1,j}}$ ($a,b=0,1$ and $\oplus$ standing for sum modulo 2). On the other hand, the information exchange between the system and ancilla will only be partial.

\section{Analysis of the quantum Fisher Information}
We make use of the formal apparatus of quantum metrology to provide a bound to the precision of the estimation of the temperature of the environment, which has been encoded in the state of the system. In particular, using the framework of quantum parameter estimation~\cite{Paris2009}, we can upper bound to said precision through the quantum Fisher Information (QFI)
\begin{equation}
	\label{eq:QFI}
	\text{QFI}_{k,j}(T,\rho_A) = \underset{\Pi_{\theta,\phi}}{\text{max}}\,F_{k,j}(\Pi_{\theta,\phi},\rho_{A_{k,j}},T),
\end{equation}
where $\Pi_{\theta,\phi}=|{\psi(\theta,\phi)}\rangle\langle{\psi(\theta,\phi)}|_{A_{k,j}}$ with $|{\psi(\theta,\phi)}\rangle_{A_{k,j}}=\cos\theta |{0}\rangle_{A_{k,j}}+e^{i\phi}\sin\theta |{1}\rangle_{A_{k,j}}$ ($\theta\in[0,\pi]$, $\phi\in[0,2\pi]$) are the elements of a positive operator valued measurement (POVM) on  ancilla $A_{k,j}$, and $F_{k,j}(\Pi_{\theta,\phi},\rho_{A_{k,j}},T)$ is the classical Fisher Information of the POVM 
\begin{equation}
	F_{k,j}(\Pi_{\theta,\phi},T,\rho_A)= \iint\,d\theta\,d\phi\, p(\theta,\phi) \left(\frac{ \partial}{ \partial T} \ln{p(\theta,\phi)} \right)^2.
\end{equation}
Here, $p(\theta,\phi)=\text{Tr}(\Pi_{\theta,\phi}\rho_{A_{k,j}})$ is the probability to get outcome $(\theta,\phi)$ from the measurement. The QFI implies the maximization of its classical counterpart over all possible POVMs, and is thus optimized over the measurement strategy on the ancillae. 
The precision of a probe in thermal equilibrium with a system is limited by the Cram\'er-Rao bound
\begin{equation}
	\left(\Delta T\right)^2 \geq \frac{1}{\text{QFI}(T,\rho)},
\end{equation}
where $(\Delta T)^2$ is the temperature variance. When evaluated over a thermal state, such lower bound to the variance of a temperature estimator (sometimes referred to as thermal Fisher information~\cite{Colisao_Scarani}) can be surpassed through the use of quantum resources such as entanglement and coherence. Quantum Fisher Information for a thermalized probe $\text{QFI}^{th} = \text{QFI}(T,\rho_S^{th})$, that is, the thermal Cram\'er-Rao bound, is \cite{Colisao_Scarani}
\begin{equation}
	\text{QFI}^{th}=\frac{1}{\bar{n}(\bar{n}+1)(2\bar{n}+1)^2}\left(\frac{\partial\bar{n}}{\partial T}\right)^2.
	\label{eq:QFI-th}
\end{equation}

\begin{figure}[t!]
\centering
\includegraphics[width=1.0\columnwidth]{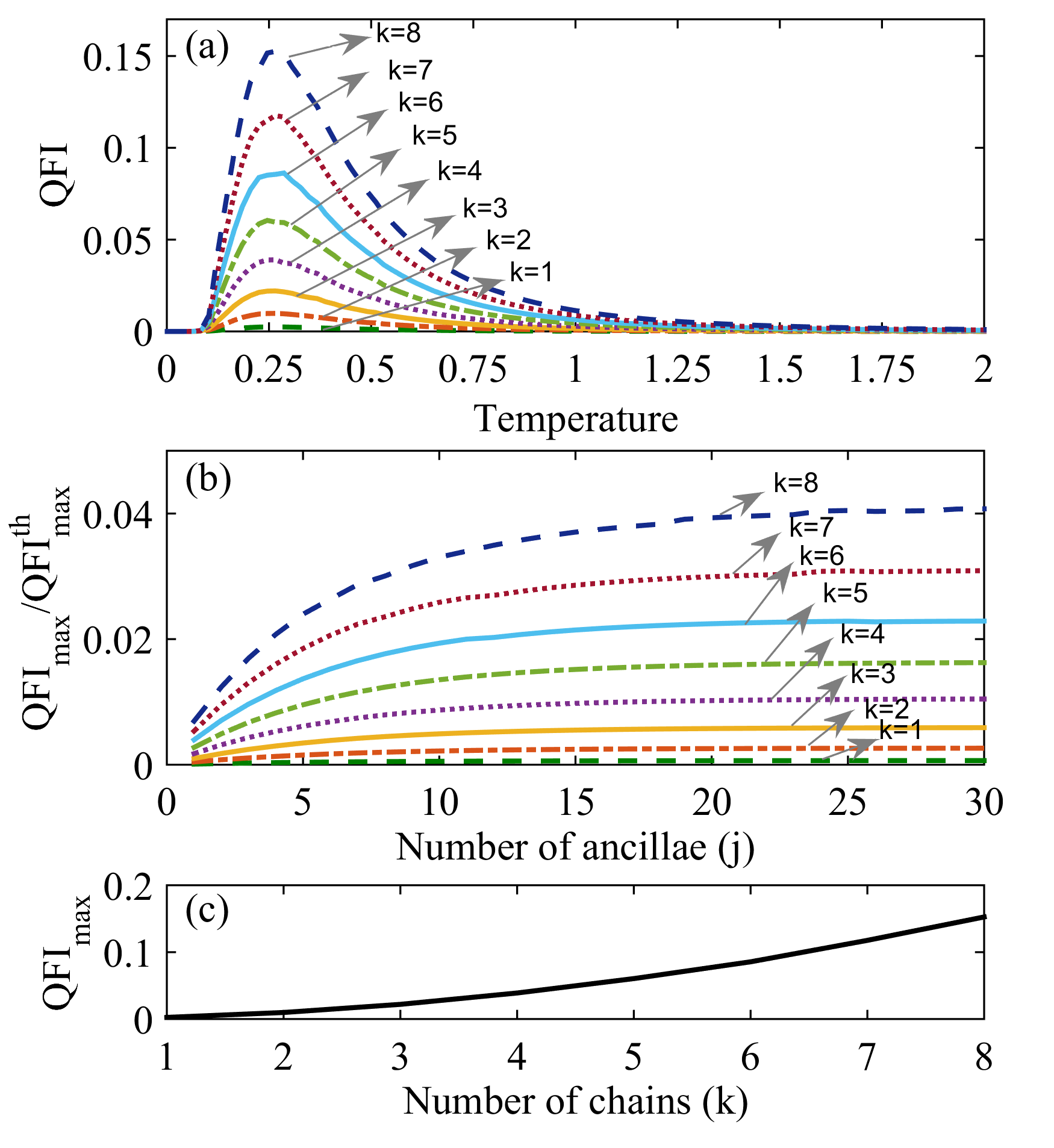}
\caption{(a) QFI \textit{versus} temperature and (b) ratio between maximum QFI and maximum QFI of the thermalized probe \textit{versus} number of ancilla on each chain in the optimized temperature and (c) maximum QFI \textit{versus} number of ancilla chains. Here we used $g\tau_{SA}\ll{\pi/2}$ and $J\tau_{A}=\pi/2$, $\text{QFI}_{max}^{th}=3.80$.}
\label{fig:QFI}
\end{figure}

The results of the maximum QFI measurement for each qubit in chain $k=1,..,8$ are shown in Fig.~\ref{fig:QFI}(a), which shows the value of the QFI for measurements performed on the ancillae of each chain $k$ as we vary the temperature. Here, we have assumed to measure the last ancilla of each chain $(j=30)$. The result of $k=1$ is equivalent to that found in the literature for the single-chain system \cite{Landi2019_Colisao}. As the number of layered chains increases, the QFI grows thus lowering the upper bound entailed by the Cram\'er-Rao bound and thus witnessing and increase in sensitivity. Fig. \ref{fig:QFI}(b) shows the ratio of maximum QFI to maximum $\text{QFI}^{th}_{max}$ as we vary the number of ancillae in each chain, i.e. varying $j$, showing that, in a sufficiently layered structure, even a few ancillae will be sufficient for our collisional thermometer to achieve high sensitivity.
Note that the value obtained for the maximum of the QFIth in the model proposed in \cite{Landi2019_Colisao} was $\text{QFI}_{max}^{th[17]}=0.015$ and in our model it was $\text{QFI}_{max}^{th}=3.80$, therefore $\text{QFI}^{th[17]}\approx 10^{-2}\text{QFI}_{max}^{th}$. Thus, even if the ratio $\text{QFI}_{max}^{[17]}/\text{QFI}_{max}^{th[17]}$ obtained in \cite{Landi2019_Colisao}  is of the same order as that obtained in our system, we will still have a sensitivity dozens of times greater for $k=8$ chains. 

Fig. \ref{fig:QFI}(c) shows the maximum value of the QFI growing with the number of layers. As we propose here a non-thermalized sensor with the system, the system-ancilla (system-sensor) interaction time is very small ($g\tau_{SA}=\pi/100$), this is done with the intention of simulating a non-thermalization between system and sensor. Any value of $g\tau_{SA}$ between $\pi/100$ and $\pi/2$ will increase the sensitivity of the sensor, that is, it will increase the QFI, at the cost of a longer thermalization time between system-sensor. The interaction time $J\tau_A=\pi/2$ was chosen so that all the information acquired by the ancilla in the system-ancilla interaction is fully transferred to the ancilla in the next chain, any value of $J\tau_A<\pi/2$ will be equivalent to a lower value of QFI.

\noindent
\section{Assessing information flow and its role in the thermometric task}
To understand the reason for the increased thermometric sensitivity, we will analyze the flow of information during the various operations carried out between system and ancillae. 

We consider the system prepared as proposed in \cite{mendonca2024InfFlow} where the authors investigated the behavior of the information flow between a main system and its environment. In that paper, a non-Markovianity measure proposed by Breuer, Laine and Piilo (BLP) \cite{Breuer2009,Breuer2010} was used to know the path taken by the information in an environment composed of qubits. The BLP measure uses the distinguishability of two initial states of a quantum system to quantify the degree of departure of a given evolution from the Markovian structure.
The lack of monotonicity of trace distance signals a backflow of information into the system stemming from the joint system-ancillae dynamics.

\begin{figure}[b!]
\centering
\includegraphics[width=1.0\columnwidth]{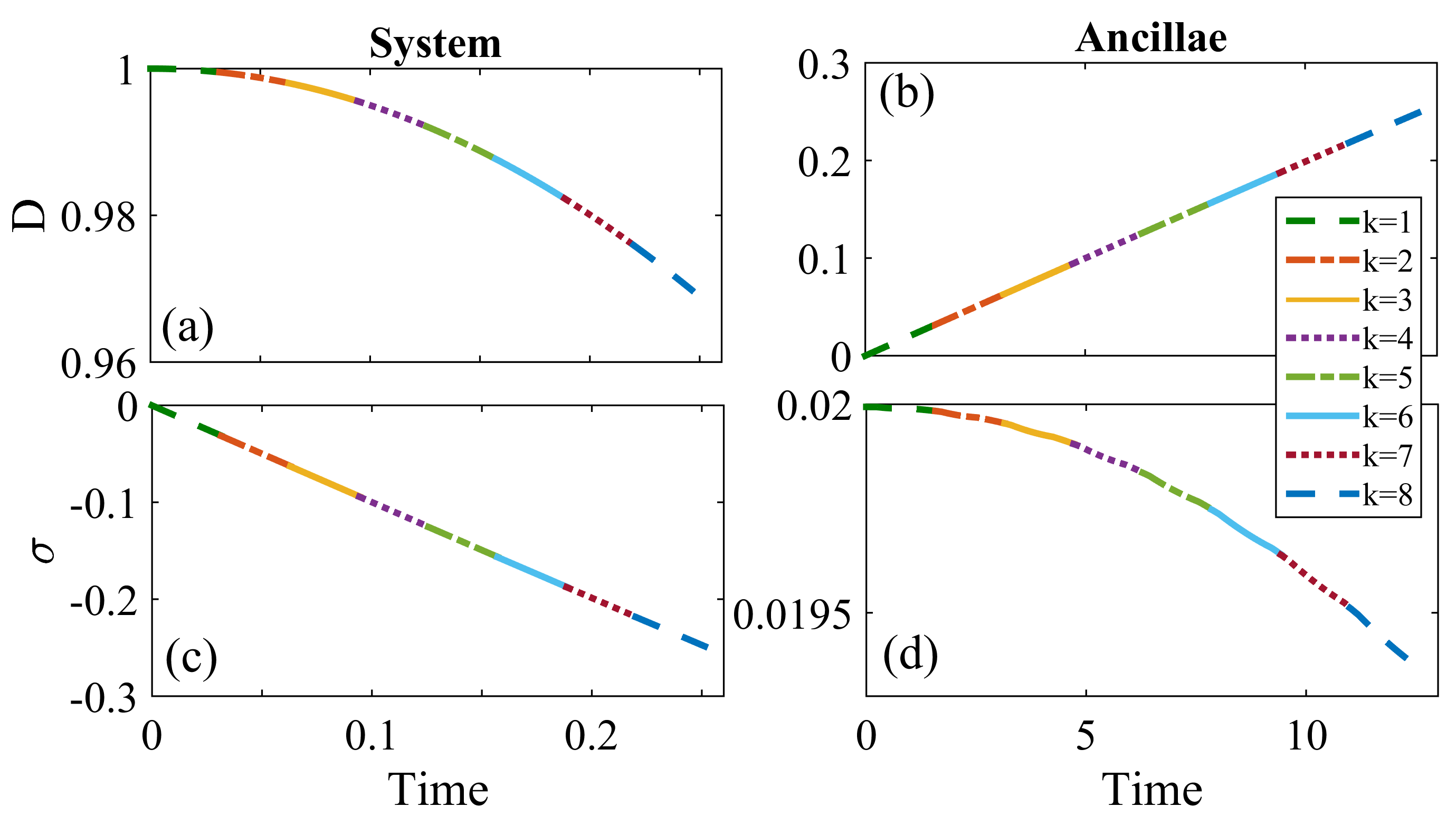}
\caption{Measure of non-Markovianity. (a) and (b) are respectively the results for the trace distance of the system and ancillae. (c) and (d) are the results for the derivative of the trace distance (measure of information flow) of the system and ancillae, respectively. Here we used $g\tau_{SA}\ll{\pi/2}$ and $J\tau_{A}=\pi/2$.}
\label{fig:Markov}
\end{figure}

\begin{figure}[t!]
\centering
    \includegraphics[width=1.0\columnwidth]{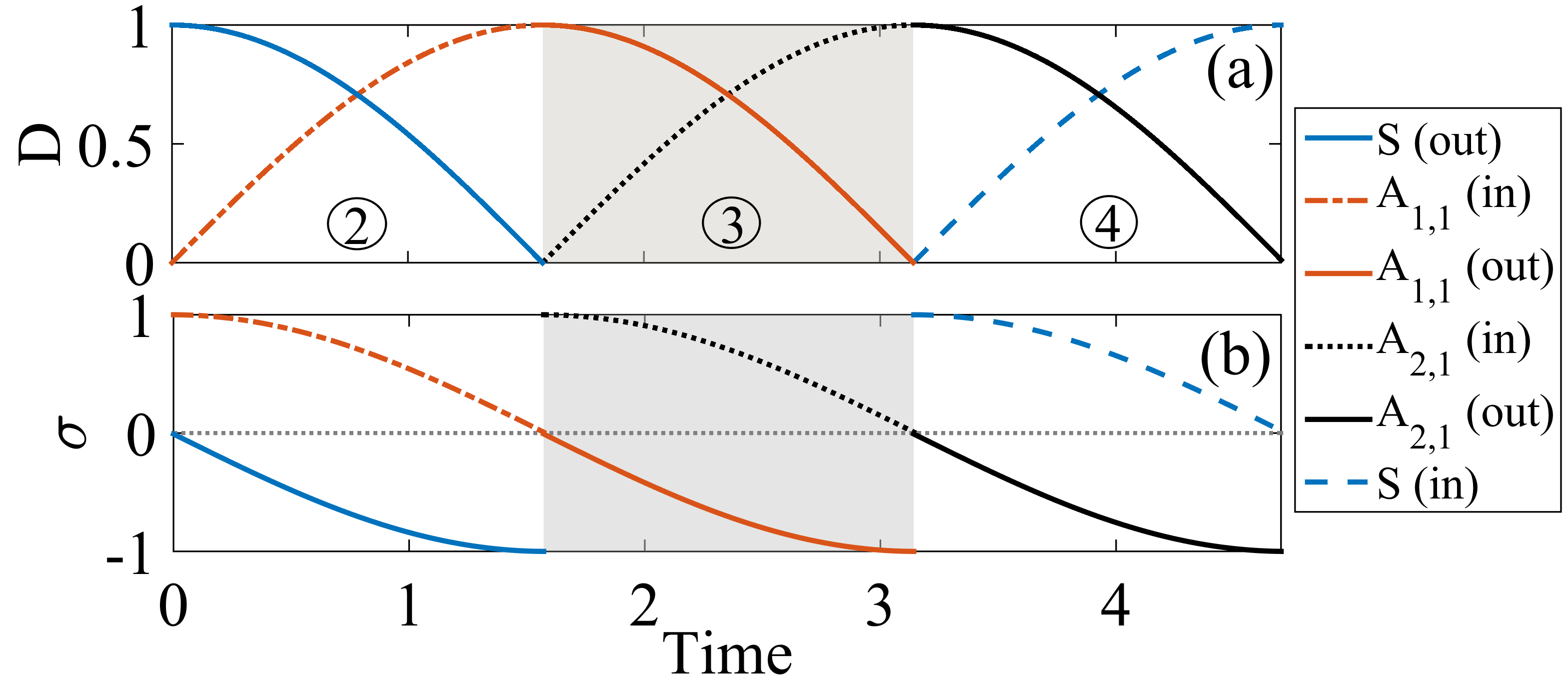}
\caption{(a) Trace distance and (b) Information flow between the qubits involved in interactions $2$, $3$ and $4$ in Fig. \ref{fig:Sistema}, the measurements are performed for an complete exchange of information between system and ancillae. Solid lines show information leaving one qubit as it arrives at another qubit, shown in dot-dashed, dashed and dotted lines. Here we used $g\tau_{SA}={\pi/2}$ and $J\tau_{A}=\pi/2$.}
\label{fig:BLP_pi2}
\end{figure}

Following Refs.~\cite{Breuer2016_colloquium,Rivas2014,Li2018,Dariusz2022}, this witnesses the non-Markovian nature of the open dynamics undergone by a qubit.
Therefore, the effect of the system-ancilla interaction can be quantified, at every instant of time, by considering the distinguishability between the the time-evolved form of such initial preparations as given by the trace distance. To study the system we will have the quantity
\begin{equation}
D\left(\rho_{S}^{(+)},\rho_{S}^{(-)};t\right)= \|\rho_{S}^{(+)}(t)-\rho_{S}^{(-)}(t)\|_{1},
\label{eq:TD_system}
\end{equation}
with $\|\cdot\|_1$ the trace distance, $\rho_{S}^{(\pm)}(t)={\rm Tr}_A\left[\rho_{S,A_{1,1}\cdots A_{N,1}}^{(\pm)}(t)\right]$ is the state of the system where ${\rm Tr}_{A}$ denotes the partial trace over all the ancillae. The results shown in this section refer to operations involving the first ancilla of each chain, that is, $A_{k,1}$.
Now, with the analysis of the ancillae, we will have
\begin{equation}
D\left(\rho_{A_{k,1}}^{(+)},\rho_{A_{k,1}}^{(-)};t\right)= \|\rho_{A_{k,1}}^{(+)}(t)-\rho_{A_{k,1}}^{(-)}(t)\|_{1},
\label{eq:TD_ancilla}
\end{equation}
where $\rho_{A_{k,1}}^{(\pm)}(t)={\rm Tr}_{S,A_{\bar{k},1}}\left[\rho_{S,A_{1,1}\cdots A_{N,j}}^{(\pm)}(t)\right]$ is the state of the ancilla $A_{k,1}$ such that ${\rm Tr}_{S,A_{\bar{k},1}}$ denotes the partial trace over all qubits except the $k^{th}$ ancilla.
To measure such dynamics, we take the initial states of the system to be $\rho_{S}^{(\pm)} (0)=\left(|{\pm}\rangle\langle{\pm}|\right)_S$ while the initial state of the ancillae is $\rho_{A_{k,j}}(0) = \bigotimes^{N}_{k=1}|{0}\rangle\langle{0}|_{k}$, with $|{\pm}\rangle=\frac{1}{2}\left(|{0}\rangle\pm|{1}\rangle\right)$. The initial state of the system-ancillae compound is the uncorrelated one $\rho_{SA}^{(\pm)}(0)=\rho_{S}^{(\pm)}(0)\otimes\rho_{A_{k,j}}(0)$. Below we will show how the information dynamics occur in our system when we consider two different interaction times between system-ancillae when $g\tau_{SA}\ll\pi/2$ (Figure~\ref{fig:Markov}) and $g\tau_{SA}=\pi/2$ (Figure~\ref{fig:BLP_pi2}).

Fig.~\ref{fig:Markov} shows the dynamics of the trace distance and its derivative of the system with the first ancilla of each chain $A_{k,1}$ for $g\tau_{SA}\ll{\pi/2}$ and $J\tau_{A}=\pi/2$. Fig.~\ref{fig:Markov} (a) and (b) report the behavior of Eq.~\eqref{eq:TD}, its derivative $\sigma(t)=\partial_t D\left(\rho_{S_{+}A_{k,1}},\rho_{S_{-}A_{k,1}};t\right)$ is represented in Fig.~\ref{fig:Markov} (c) and (d).
Fig. \ref{fig:Markov}(a) and (c) refer to measurements on the system, i.e., tracing the ancillae while (b) and (d) show the measurements on the ancillae $A_{k,1}$ while we trace the system.
We observe a flow of information between system and ancillae, growing with the number of chains being considered.
We know that when $\sigma<0$ there is information leaving a given system, just as there is information entering when $\sigma>0$ \cite{Breuer2009, Breuer2010,mendonca2024InfFlow}. In Figure \ref{fig:Markov}(c) we observe $\sigma<0$ indicating that information is leaving the qubit system at the same time that there is information arriving at the ancillae [$\sigma>0$ in Figure \ref{fig:Markov}(d)]. Therefore, we easily conclude that information is flowing from the qubit system to the ancillae.

Knowing the dynamics of the information flow, we ask ourselves how it might be influencing the sensitivity gain of our thermometer. So, in order to gather insight in the dynamics of information flow during the thermometric task, in the following numerical experiments we will consider $g\tau_{SA}=J\tau_{A}=\pi/2$, i.e., a full exchange of information between system-ancilla and ancilla-ancilla.
Measurements were performed on all parties involved in the processes ruled by $V_{S,A_{k,j}}$ and $V_{A_{k,j},A_{k+1,j}}$. However, for a clearer visualization we plot only the results for $j=1$ and $k=1,2$, i.e. 
$V_{S,A_{1,1}}$ (measuring $S$ and $A_{1,1}$), $V_{A_{1,1},A_{2,1}}$ (measuring $A_{1,1}$ and $A_{2,1}$) and $V_{S,A_{2,1}}$ (measuring $S$ and $A_{2,1}$). These are the operations
illustrated in steps $2$, $3$ and $4$ of Fig. \ref{fig:Sistema}(a). These analyses and results does not depend on the choice of $j$ made here.

\begin{figure}[b!]
\centering    \includegraphics[width=1.0\columnwidth]{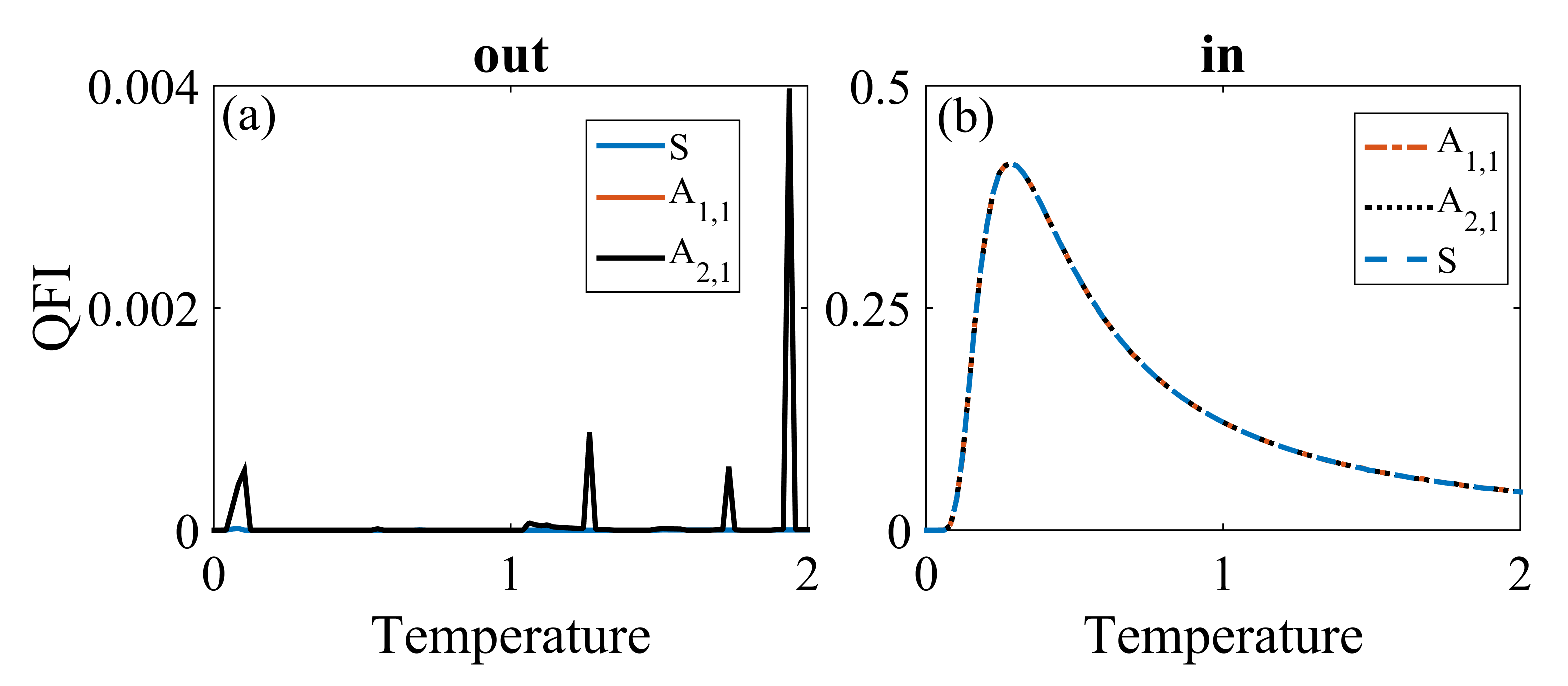}
\caption{QFI performed on qubits involved in interactions $2$, $3$ and $4$ in Fig. \ref{fig:Sistema}. We consider a full exchange of information between system and ancillae. Measurements are performed on qubits when information is (a) leaving and (b) entering. Here we used $g\tau_{SA}={\pi/2}$ and $J\tau_{A}=\pi/2$.}
	\label{fig:Fluxo_pi2}
\end{figure}

We show the behavior of the trace distance, its derivative $\sigma$ and the QFI in Fig.~\ref{fig:BLP_pi2} and Fig.~\ref{fig:Fluxo_pi2}, respectively. We use the protocol put forward in  Ref.~\cite{mendonca2024InfFlow}, where such tools have been used to analyze the flow of information passing through a qubit environment.

Fig.~\ref{fig:BLP_pi2}(a) and Fig.~\ref{fig:BLP_pi2}(b) show the dynamics of the trace distance and its derivative, respectively, for the states described above. 
Fig.~\ref{fig:BLP_pi2}(b) shows clearly how each collision (either system-ancilla or ancilla-ancilla) results in a pouring of information that flows according to the preferential direction
\begin{equation}
S\to\text{chain \textit{k}}\to\text{chain \textit{k}+1}\to S
\label{percurso_S-A}
\end{equation}
for $k=1,...,N$ since the positions of the ancillae such that $k$ is odd and $k+1$ is even. Here we can also see that information leaves a given qubit [as witnessed by having $\sigma<0$, cf. solid line in Fig.~\ref{fig:BLP_pi2}(b)], and reaches the other qubit involved in the operation [resulting in $\sigma>0$, cf. dot-dashed, dotted, and dashed lines]. 

We now evaluate the QFI using the same protocol described above, i.e., we measure the QFI on each qubit involved in the operations when we use $g\tau_{SA}=\pi/2$ and $J\tau_{A}=\pi/2$. Since the system qubit acquires information about the environment and transmits such information to the ancillae through exchange interactions in such a way that the relation \eqref{percurso_S-A} is the information path. Fig.~\ref{fig:Fluxo_pi2}(a) shows that the QFI cannot be measured at qubits that transmit the information, we highlight that the QFI measurement was performed at the end of the operation where all the information is being transferred to the next qubit. 
Therefore, Fig.~\ref{fig:Fluxo_pi2}(b) shows that the same QFI can be extracted from measurements performed on qubits that are receiving the information during each operation.
We see that the QFI measurement does not vary when $g\tau_{SA}=J\tau_{A}=\pi/2$ there is no gain in sensor sensitivity, this happens because the same amount of information is being transferred in the full Swap operation [cf. Fig.~\ref{fig:BLP_pi2}(b)].

\begin{figure}[ht!]
\centering
    \includegraphics[width=0.9\columnwidth]{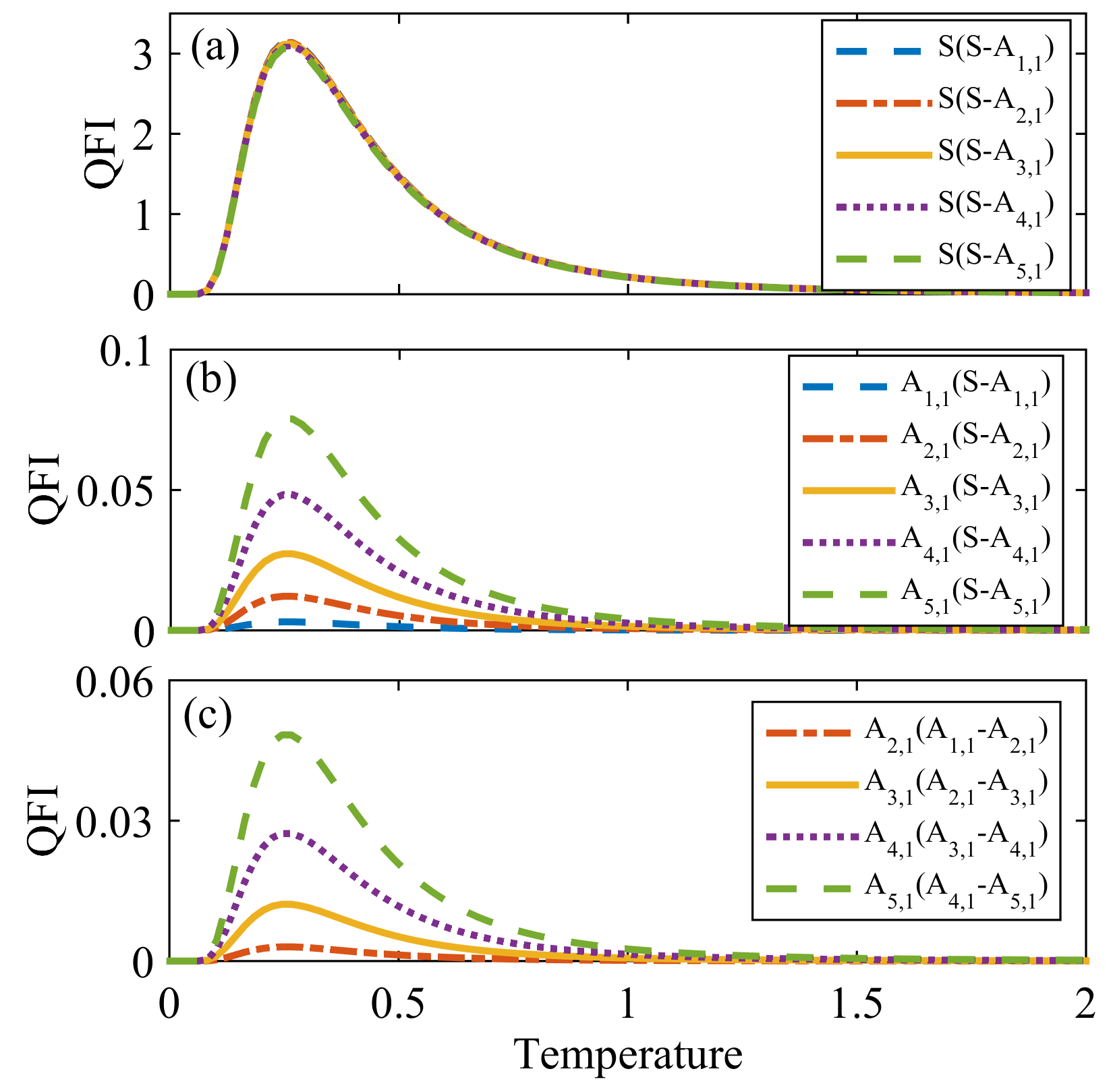}
\caption{QFI resulting from an incomplete exchange of information between system and ancillae. The measurements needed to perform the estimation are performed on (a) the system (during the system-ancilla interactions); (b) the ancillae (during the system-ancilla interactions); (c) the ancillae (during the ancilla-ancilla interactions). Here we used $g\tau_{SA}\ll{\pi/2}$ and $J\tau_{A}=\pi/2$.}
\label{fig:Fluxo_pi100}
\end{figure}

We now extend the analysis and return to the case of $g\tau_{SA}\ll\pi/2$, i.e. an incomplete exchange of information between system and ancillae. In Fig. \ref{fig:Fluxo_pi100} we consider the results corresponding to $j=1$ and a compound comprising 5 chains. The QFI for the system and the ancillae are shown in Figs. \ref{fig:Fluxo_pi100}(a) and (b), respectively.
Fig.~\ref{fig:Fluxo_pi100}(c) shows the QFI for the ancilla that receives information during the ancilla-ancilla interaction. Since the interactions characterizes an full Swap operation, i.e. $J\tau_{A}=\pi/2$, the QFI measurements performed on the ancillae from which the information is leaving the qubit follow the result shown in Fig.~\ref{fig:Fluxo_pi2}(a), thus we will not repeat such results here. 
As in Fig. \ref{fig:Fluxo_pi2}(b), Fig. \ref{fig:Fluxo_pi100}(a) shows that the same amount of information can be measured in the different parts of the compound. However,  Figs.~\ref{fig:Fluxo_pi100}(b) and (c) show that there is an increase in the sensitivity of our thermometer as we increase the number of ancillary chains. Therefore, this represents there is an accumulation of information caused by the short interaction time between system and ancillae $\tau_{SA}$, as is evident in the characterization of the distinguishability in Fig. \ref{fig:Markov}(d). In short, since we can only obtain the QFI when the information flows into an element, we can state that there is also information returning to the system leading to an accumulation of information that is reflected in the gain in sensitivity of the thermometric sensor.

\begin{figure}[b!]
\centering \includegraphics[width=0.9\columnwidth]{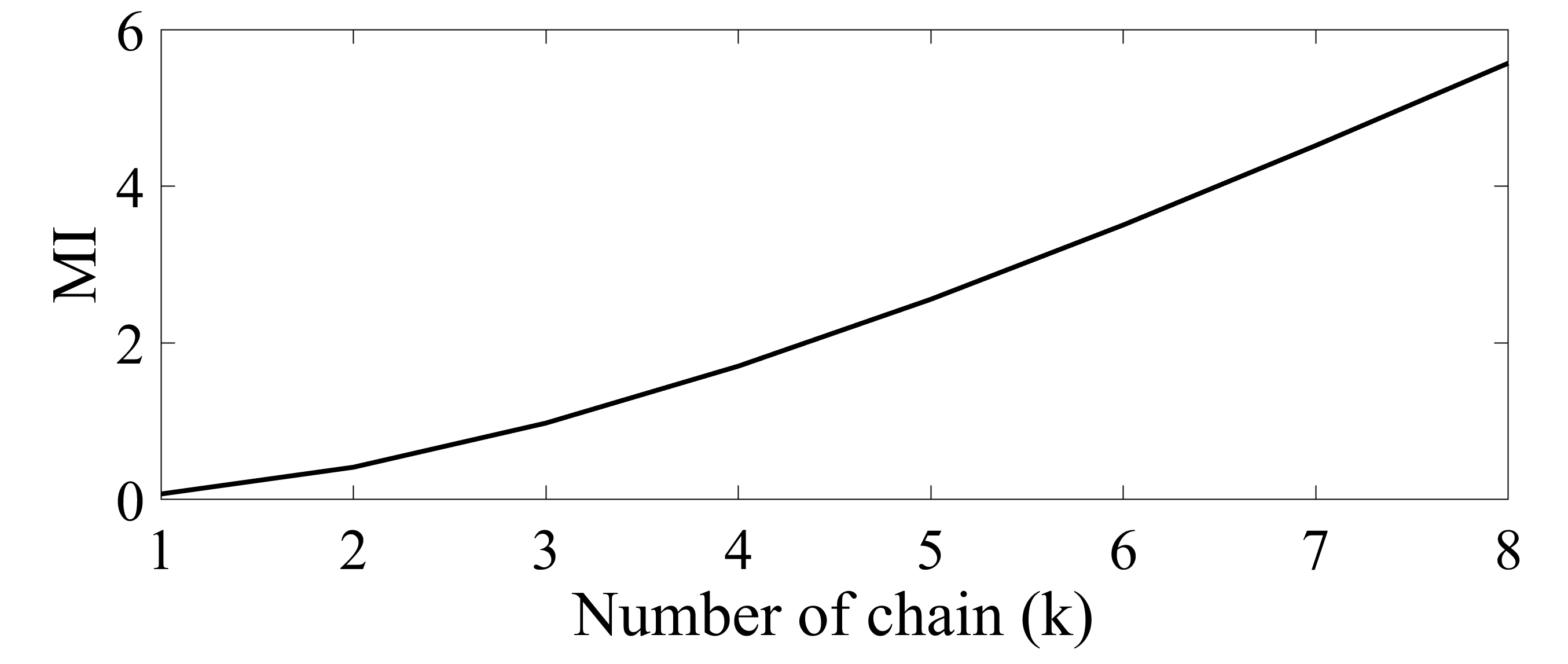}
\caption{Mutual information between the system and the qubits of the ancilla chains during operations $V_{S,A_{k,n}}$. Here we used $g\tau_{SA}\ll{\pi/2}$ and $J\tau_{A}=\pi/2$.}
\label{fig:IM}
\end{figure}

The accumulation of information can also be seen when we calculate the mutual information between $S$ and an ancillary chain. We compute the mutual information between the elements with $j=n$ of the $k^\text{th}$ chain in all interactions as
\begin{equation}
I\left(\rho_{S}:\rho_{A_{k,n}}\right)= \mathcal{E}(\rho_S)+\mathcal{E}(\rho_{A_{k,n}})-\mathcal{E}(\rho_{SA_{k,n}}).
\end{equation}
where $\mathcal{E}(\rho)= -\text{tr}(\rho \text{log}\rho)$ is the von Neumann entropy \cite{NielsenLivro2010}. Fig. \ref{fig:IM} shows that as we increase the number of ancilla chains, the greater the amount of information shared between the system and the ancillae.

\noindent
\section{Conclusion}
We have built a model to improve and increase the sensitivity of a quantum thermometer based on the use of a multilayered system. Even with a short interaction time, i.e. without requesting thermalization of the system with the thermometer, we can adjust the sensitivity of the latter by increasing the number of its ancillary chains.
Our model has been shown to be effective when we use only one ancilla in each chain, i.e. $j=1$, as long as we put a sufficient number of chains $k$. However, although this means fewer iterations, any increase in the number of chains $k$ will result in an exponential increase in the Hilbert space of the system, computationally limiting the construction of the sensor.
We have unveiled a tight link between the sensitivity of the estimation of temperature as  provided by the QFI and the information flow among various parts of the compound. The accumulation of information entailed by the information flow leads to better sensitivity (i.e. a larger QFI). This  can also be seen in the increase of mutual information with the number of chains comprised in the thermometer.
Note, however, that the relationship between information flow and the improvement of a sensor's sensitivity also depends on the type of interaction between the system and the probe. To understand whether this connection is valid, it would be necessary to perform tests on various sensor models with different interaction dynamics. Another point that can also be worked on in the improvement of sensors is the use of other quantum resources such as coherence and entanglement. In our sensor, we measured both, but we did not find the creation of coherence in the dynamics of the ancillae qubits. As for entanglement, we measured the quantum negativity, which, although its maximum increases in the same way as the maximum of the QFI, has a small influence on the dynamics, having a value of 0.065QFI$_{max}$.


\section*{Acknowledgments}
The authors thank discussions with G. T. Landi.
TMM thanks the grants No. 2021/01277-2 and No. 2022/09219-4, S\~{a}o Paulo Research Foundation (FAPESP).
DOSP acknowledges the support by the Brazilian funding agencies CNPq (Grant No. 304891/2022-3), FAPESP (Grant No. 2017/03727-0), and the Brazilian National Institute of Science and Technology of Quantum Information (INCT/IQ).
M.P. acknowledges the support by the European Union’s Horizon Europe EIC Pathfinder project QuCoM (Grant Agreement No.101046973), the Leverhulme Trust Grant UltraQuTe (Grant No. RGP-2018-266), the Royal Society Wolfson Fellowship (RSWF/R3/183013), the U.K. EPSRC (EP/T028424/1), the Department for the Economy Northern Ireland under the U.S.-Ireland R\&D Partnership Programme, the "Italian National Quantum Science and Technology Institute (NQSTI)" (PE0000023)-SPOKE 2 through project ASpEQCt, and the National Centre for HPC, Big Data and Quantum Computing (HPC) (CN00000013)-SPOKE 10 through project HyQELM. 

\bibliography{referencias}

\begin{thebibliography}{29}%
\makeatletter
\providecommand \@ifxundefined [1]{%
 \@ifx{#1\undefined}
}%
\providecommand \@ifnum [1]{%
 \ifnum #1\expandafter \@firstoftwo
 \else \expandafter \@secondoftwo
 \fi
}%
\providecommand \@ifx [1]{%
 \ifx #1\expandafter \@firstoftwo
 \else \expandafter \@secondoftwo
 \fi
}%
\providecommand \natexlab [1]{#1}%
\providecommand \enquote  [1]{``#1''}%
\providecommand \bibnamefont  [1]{#1}%
\providecommand \bibfnamefont [1]{#1}%
\providecommand \citenamefont [1]{#1}%
\providecommand \href@noop [0]{\@secondoftwo}%
\providecommand \href [0]{\begingroup \@sanitize@url \@href}%
\providecommand \@href[1]{\@@startlink{#1}\@@href}%
\providecommand \@@href[1]{\endgroup#1\@@endlink}%
\providecommand \@sanitize@url [0]{\catcode `\\12\catcode `\$12\catcode
  `\&12\catcode `\#12\catcode `\^12\catcode `\_12\catcode `\%12\relax}%
\providecommand \@@startlink[1]{}%
\providecommand \@@endlink[0]{}%
\providecommand \url  [0]{\begingroup\@sanitize@url \@url }%
\providecommand \@url [1]{\endgroup\@href {#1}{\urlprefix }}%
\providecommand \urlprefix  [0]{URL }%
\providecommand \Eprint [0]{\href }%
\providecommand \doibase [0]{https://doi.org/}%
\providecommand \selectlanguage [0]{\@gobble}%
\providecommand \bibinfo  [0]{\@secondoftwo}%
\providecommand \bibfield  [0]{\@secondoftwo}%
\providecommand \translation [1]{[#1]}%
\providecommand \BibitemOpen [0]{}%
\providecommand \bibitemStop [0]{}%
\providecommand \bibitemNoStop [0]{.\EOS\space}%
\providecommand \EOS [0]{\spacefactor3000\relax}%
\providecommand \BibitemShut  [1]{\csname bibitem#1\endcsname}%
\let\auto@bib@innerbib\@empty
\bibitem [{\citenamefont {Giovannetti}\ \emph {et~al.}(2006)\citenamefont
  {Giovannetti}, \citenamefont {Lloyd},\ and\ \citenamefont
  {Maccone}}]{Giovannetti2006}%
  \BibitemOpen
  \bibfield  {author} {\bibinfo {author} {\bibfnamefont {V.}~\bibnamefont
  {Giovannetti}}, \bibinfo {author} {\bibfnamefont {S.}~\bibnamefont {Lloyd}},\
  and\ \bibinfo {author} {\bibfnamefont {L.}~\bibnamefont {Maccone}},\
  }\bibfield  {title} {\bibinfo {title} {Quantum metrology},\ }\href
  {https://doi.org/10.1103/PhysRevLett.96.010401} {\bibfield  {journal}
  {\bibinfo  {journal} {Phys. Rev. Lett.}\ }\textbf {\bibinfo {volume} {96}},\
  \bibinfo {pages} {010401} (\bibinfo {year} {2006})}\BibitemShut {NoStop}%
\bibitem [{\citenamefont {Maccone}(2013)}]{Maccone2013}%
  \BibitemOpen
  \bibfield  {author} {\bibinfo {author} {\bibfnamefont {L.}~\bibnamefont
  {Maccone}},\ }\bibfield  {title} {\bibinfo {title} {Intuitive reason for the
  usefulness of entanglement in quantum metrology},\ }\href
  {https://doi.org/10.1103/PhysRevA.88.042109} {\bibfield  {journal} {\bibinfo
  {journal} {Phys. Rev. A}\ }\textbf {\bibinfo {volume} {88}},\ \bibinfo
  {pages} {042109} (\bibinfo {year} {2013})}\BibitemShut {NoStop}%
\bibitem [{\citenamefont {Huang}\ \emph {et~al.}(2016)\citenamefont {Huang},
  \citenamefont {Macchiavello},\ and\ \citenamefont {Maccone}}]{Maccone2016}%
  \BibitemOpen
  \bibfield  {author} {\bibinfo {author} {\bibfnamefont {Z.}~\bibnamefont
  {Huang}}, \bibinfo {author} {\bibfnamefont {C.}~\bibnamefont
  {Macchiavello}},\ and\ \bibinfo {author} {\bibfnamefont {L.}~\bibnamefont
  {Maccone}},\ }\bibfield  {title} {\bibinfo {title} {Usefulness of
  entanglement-assisted quantum metrology},\ }\href
  {https://doi.org/10.1103/PhysRevA.94.012101} {\bibfield  {journal} {\bibinfo
  {journal} {Phys. Rev. A}\ }\textbf {\bibinfo {volume} {94}},\ \bibinfo
  {pages} {012101} (\bibinfo {year} {2016})}\BibitemShut {NoStop}%
\bibitem [{\citenamefont {Micadei}\ \emph {et~al.}(2015)\citenamefont
  {Micadei}, \citenamefont {Rowlands}, \citenamefont {Pollock}, \citenamefont
  {C{\'{e}}leri}, \citenamefont {Serra},\ and\ \citenamefont
  {Modi}}]{Serra2015a}%
  \BibitemOpen
  \bibfield  {author} {\bibinfo {author} {\bibfnamefont {K.}~\bibnamefont
  {Micadei}}, \bibinfo {author} {\bibfnamefont {D.~A.}\ \bibnamefont
  {Rowlands}}, \bibinfo {author} {\bibfnamefont {F.~A.}\ \bibnamefont
  {Pollock}}, \bibinfo {author} {\bibfnamefont {L.~C.}\ \bibnamefont
  {C{\'{e}}leri}}, \bibinfo {author} {\bibfnamefont {R.~M.}\ \bibnamefont
  {Serra}},\ and\ \bibinfo {author} {\bibfnamefont {K.}~\bibnamefont {Modi}},\
  }\bibfield  {title} {\bibinfo {title} {Coherent measurements in quantum
  metrology},\ }\href {https://doi.org/10.1088/1367-2630/17/2/023057}
  {\bibfield  {journal} {\bibinfo  {journal} {New Journal of Physics}\ }\textbf
  {\bibinfo {volume} {17}},\ \bibinfo {pages} {023057} (\bibinfo {year}
  {2015})}\BibitemShut {NoStop}%
\bibitem [{\citenamefont {Pires}\ \emph {et~al.}(2018)\citenamefont {Pires},
  \citenamefont {Silva}, \citenamefont {deAzevedo}, \citenamefont
  {Soares-Pinto},\ and\ \citenamefont {Filgueiras}}]{Pires2018}%
  \BibitemOpen
  \bibfield  {author} {\bibinfo {author} {\bibfnamefont {D.~P.}\ \bibnamefont
  {Pires}}, \bibinfo {author} {\bibfnamefont {I.~A.}\ \bibnamefont {Silva}},
  \bibinfo {author} {\bibfnamefont {E.~R.}\ \bibnamefont {deAzevedo}}, \bibinfo
  {author} {\bibfnamefont {D.~O.}\ \bibnamefont {Soares-Pinto}},\ and\ \bibinfo
  {author} {\bibfnamefont {J.~G.}\ \bibnamefont {Filgueiras}},\ }\bibfield
  {title} {\bibinfo {title} {Coherence orders, decoherence, and quantum
  metrology},\ }\href {https://doi.org/10.1103/PhysRevA.98.032101} {\bibfield
  {journal} {\bibinfo  {journal} {Phys. Rev. A}\ }\textbf {\bibinfo {volume}
  {98}},\ \bibinfo {pages} {032101} (\bibinfo {year} {2018})}\BibitemShut
  {NoStop}%
\bibitem [{\citenamefont {Castellini}\ \emph {et~al.}(2019)\citenamefont
  {Castellini}, \citenamefont {Lo~Franco}, \citenamefont {Lami}, \citenamefont
  {Winter}, \citenamefont {Adesso},\ and\ \citenamefont
  {Compagno}}]{Castellini2019}%
  \BibitemOpen
  \bibfield  {author} {\bibinfo {author} {\bibfnamefont {A.}~\bibnamefont
  {Castellini}}, \bibinfo {author} {\bibfnamefont {R.}~\bibnamefont
  {Lo~Franco}}, \bibinfo {author} {\bibfnamefont {L.}~\bibnamefont {Lami}},
  \bibinfo {author} {\bibfnamefont {A.}~\bibnamefont {Winter}}, \bibinfo
  {author} {\bibfnamefont {G.}~\bibnamefont {Adesso}},\ and\ \bibinfo {author}
  {\bibfnamefont {G.}~\bibnamefont {Compagno}},\ }\bibfield  {title} {\bibinfo
  {title} {Indistinguishability-enabled coherence for quantum metrology},\
  }\href {https://doi.org/10.1103/PhysRevA.100.012308} {\bibfield  {journal}
  {\bibinfo  {journal} {Phys. Rev. A}\ }\textbf {\bibinfo {volume} {100}},\
  \bibinfo {pages} {012308} (\bibinfo {year} {2019})}\BibitemShut {NoStop}%
\bibitem [{\citenamefont {Maccone}\ and\ \citenamefont
  {Riccardi}(2020)}]{Maccone2020}%
  \BibitemOpen
  \bibfield  {author} {\bibinfo {author} {\bibfnamefont {L.}~\bibnamefont
  {Maccone}}\ and\ \bibinfo {author} {\bibfnamefont {A.}~\bibnamefont
  {Riccardi}},\ }\bibfield  {title} {\bibinfo {title} {Squeezing metrology: a
  unified framework},\ }\href {https://doi.org/10.22331/q-2020-07-09-292}
  {\bibfield  {journal} {\bibinfo  {journal} {Quantum}\ }\textbf {\bibinfo
  {volume} {4}},\ \bibinfo {pages} {292} (\bibinfo {year} {2020})}\BibitemShut
  {NoStop}%
\bibitem [{\citenamefont {De~Pasquale}\ and\ \citenamefont
  {Stace}(2018)}]{DePasquale2018}%
  \BibitemOpen
  \bibfield  {author} {\bibinfo {author} {\bibfnamefont {A.}~\bibnamefont
  {De~Pasquale}}\ and\ \bibinfo {author} {\bibfnamefont {T.~M.}\ \bibnamefont
  {Stace}},\ }\bibinfo {title} {Quantum thermometry},\ in\ \href
  {https://doi.org/10.1007/978-3-319-99046-0_21} {\emph {\bibinfo {booktitle}
  {Thermodynamics in the Quantum Regime: Fundamental Aspects and New
  Directions}}},\ \bibinfo {editor} {edited by\ \bibinfo {editor}
  {\bibfnamefont {F.}~\bibnamefont {Binder}}, \bibinfo {editor} {\bibfnamefont
  {L.~A.}\ \bibnamefont {Correa}}, \bibinfo {editor} {\bibfnamefont
  {C.}~\bibnamefont {Gogolin}}, \bibinfo {editor} {\bibfnamefont
  {J.}~\bibnamefont {Anders}},\ and\ \bibinfo {editor} {\bibfnamefont
  {G.}~\bibnamefont {Adesso}}}\ (\bibinfo  {publisher} {Springer International
  Publishing},\ \bibinfo {address} {Cham},\ \bibinfo {year} {2018})\ pp.\
  \bibinfo {pages} {503--527}\BibitemShut {NoStop}%
\bibitem [{\citenamefont {Ciccarello}\ \emph {et~al.}(2022)\citenamefont
  {Ciccarello}, \citenamefont {Lorenzo}, \citenamefont {Giovannetti},\ and\
  \citenamefont {Palma}}]{CICCARELLO2021}%
  \BibitemOpen
  \bibfield  {author} {\bibinfo {author} {\bibfnamefont {F.}~\bibnamefont
  {Ciccarello}}, \bibinfo {author} {\bibfnamefont {S.}~\bibnamefont {Lorenzo}},
  \bibinfo {author} {\bibfnamefont {V.}~\bibnamefont {Giovannetti}},\ and\
  \bibinfo {author} {\bibfnamefont {G.~M.}\ \bibnamefont {Palma}},\ }\bibfield
  {title} {\bibinfo {title} {Quantum collision models: Open system dynamics
  from repeated interactions},\ }\href
  {https://doi.org/https://doi.org/10.1016/j.physrep.2022.01.001} {\bibfield
  {journal} {\bibinfo  {journal} {Physics Reports}\ }\textbf {\bibinfo {volume}
  {954}},\ \bibinfo {pages} {1} (\bibinfo {year} {2022})}\BibitemShut {NoStop}%
\bibitem [{\citenamefont {Ciccarello}(2017)}]{Ciccarello2017}%
  \BibitemOpen
  \bibfield  {author} {\bibinfo {author} {\bibfnamefont {F.}~\bibnamefont
  {Ciccarello}},\ }\bibfield  {title} {\bibinfo {title} {Collision models in
  quantum optics},\ }\href@noop {} {\bibfield  {journal} {\bibinfo  {journal}
  {Quantum Measurements and Quantum Metrology}\ }\textbf {\bibinfo {volume}
  {4}},\ \bibinfo {pages} {53} (\bibinfo {year} {2017})}\BibitemShut {NoStop}%
\bibitem [{\citenamefont {Rodrigues}\ \emph {et~al.}(2019)\citenamefont
  {Rodrigues}, \citenamefont {De~Chiara}, \citenamefont {Paternostro},\ and\
  \citenamefont {Landi}}]{Paternostro2019b}%
  \BibitemOpen
  \bibfield  {author} {\bibinfo {author} {\bibfnamefont {F.~L.~S.}\
  \bibnamefont {Rodrigues}}, \bibinfo {author} {\bibfnamefont {G.}~\bibnamefont
  {De~Chiara}}, \bibinfo {author} {\bibfnamefont {M.}~\bibnamefont
  {Paternostro}},\ and\ \bibinfo {author} {\bibfnamefont {G.~T.}\ \bibnamefont
  {Landi}},\ }\bibfield  {title} {\bibinfo {title} {Thermodynamics of weakly
  coherent collisional models},\ }\href
  {https://doi.org/10.1103/PhysRevLett.123.140601} {\bibfield  {journal}
  {\bibinfo  {journal} {Phys. Rev. Lett.}\ }\textbf {\bibinfo {volume} {123}},\
  \bibinfo {pages} {140601} (\bibinfo {year} {2019})}\BibitemShut {NoStop}%
\bibitem [{\citenamefont {Filippov}(2022)}]{Colisao_Filippov2022}%
  \BibitemOpen
  \bibfield  {author} {\bibinfo {author} {\bibfnamefont {S.}~\bibnamefont
  {Filippov}},\ }\bibfield  {title} {\bibinfo {title} {Multipartite
  correlations in quantum collision models},\ }\bibfield  {journal} {\bibinfo
  {journal} {Entropy}\ }\textbf {\bibinfo {volume} {24}},\ \href
  {https://doi.org/10.3390/e24040508} {10.3390/e24040508} (\bibinfo {year}
  {2022})\BibitemShut {NoStop}%
\bibitem [{\citenamefont {Landi}(2021)}]{ColisaoBateria_Landi2021}%
  \BibitemOpen
  \bibfield  {author} {\bibinfo {author} {\bibfnamefont {G.~T.}\ \bibnamefont
  {Landi}},\ }\bibfield  {title} {\bibinfo {title} {Battery charging in
  collision models with bayesian risk strategies},\ }\bibfield  {journal}
  {\bibinfo  {journal} {Entropy}\ }\textbf {\bibinfo {volume} {23}},\ \href
  {https://doi.org/10.3390/e23121627} {10.3390/e23121627} (\bibinfo {year}
  {2021})\BibitemShut {NoStop}%
\bibitem [{\citenamefont {McCloskey}\ and\ \citenamefont
  {Paternostro}(2014)}]{Paternostro2014c}%
  \BibitemOpen
  \bibfield  {author} {\bibinfo {author} {\bibfnamefont {R.}~\bibnamefont
  {McCloskey}}\ and\ \bibinfo {author} {\bibfnamefont {M.}~\bibnamefont
  {Paternostro}},\ }\bibfield  {title} {\bibinfo {title} {Non-markovianity and
  system-environment correlations in a microscopic collision model},\ }\href
  {https://doi.org/10.1103/PhysRevA.89.052120} {\bibfield  {journal} {\bibinfo
  {journal} {Phys. Rev. A}\ }\textbf {\bibinfo {volume} {89}},\ \bibinfo
  {pages} {052120} (\bibinfo {year} {2014})}\BibitemShut {NoStop}%
\bibitem [{\citenamefont {O’Connor}\ \emph {et~al.}(2021)\citenamefont
  {O’Connor}, \citenamefont {Vacchini},\ and\ \citenamefont
  {Campbell}}]{ColisaoEstoc_Steve2021}%
  \BibitemOpen
  \bibfield  {author} {\bibinfo {author} {\bibfnamefont {E.}~\bibnamefont
  {O’Connor}}, \bibinfo {author} {\bibfnamefont {B.}~\bibnamefont
  {Vacchini}},\ and\ \bibinfo {author} {\bibfnamefont {S.}~\bibnamefont
  {Campbell}},\ }\bibfield  {title} {\bibinfo {title} {Stochastic collisional
  quantum thermometry},\ }\bibfield  {journal} {\bibinfo  {journal} {Entropy}\
  }\textbf {\bibinfo {volume} {23}},\ \href {https://doi.org/10.3390/e23121634}
  {10.3390/e23121634} (\bibinfo {year} {2021})\BibitemShut {NoStop}%
\bibitem [{\citenamefont {Bouton}\ \emph {et~al.}(2020)\citenamefont {Bouton},
  \citenamefont {Nettersheim}, \citenamefont {Burgardt}, \citenamefont {Adam},
  \citenamefont {Lutz},\ and\ \citenamefont {Widera}}]{Lutz2020}%
  \BibitemOpen
  \bibfield  {author} {\bibinfo {author} {\bibfnamefont {Q.}~\bibnamefont
  {Bouton}}, \bibinfo {author} {\bibfnamefont {J.}~\bibnamefont {Nettersheim}},
  \bibinfo {author} {\bibfnamefont {S.}~\bibnamefont {Burgardt}}, \bibinfo
  {author} {\bibfnamefont {D.}~\bibnamefont {Adam}}, \bibinfo {author}
  {\bibfnamefont {E.}~\bibnamefont {Lutz}},\ and\ \bibinfo {author}
  {\bibfnamefont {A.}~\bibnamefont {Widera}},\ }\href@noop {} {\bibinfo {title}
  {An endoreversible quantum heat engine driven by atomic collisions}}
  (\bibinfo {year} {2020}),\ \Eprint {https://arxiv.org/abs/2009.10946}
  {arXiv:2009.10946 [quant-ph]} \BibitemShut {NoStop}%
\bibitem [{\citenamefont {Seah}\ \emph {et~al.}(2019)\citenamefont {Seah},
  \citenamefont {Nimmrichter}, \citenamefont {Grimmer}, \citenamefont {Santos},
  \citenamefont {Scarani},\ and\ \citenamefont {Landi}}]{Landi2019_Colisao}%
  \BibitemOpen
  \bibfield  {author} {\bibinfo {author} {\bibfnamefont {S.}~\bibnamefont
  {Seah}}, \bibinfo {author} {\bibfnamefont {S.}~\bibnamefont {Nimmrichter}},
  \bibinfo {author} {\bibfnamefont {D.}~\bibnamefont {Grimmer}}, \bibinfo
  {author} {\bibfnamefont {J.~P.}\ \bibnamefont {Santos}}, \bibinfo {author}
  {\bibfnamefont {V.}~\bibnamefont {Scarani}},\ and\ \bibinfo {author}
  {\bibfnamefont {G.~T.}\ \bibnamefont {Landi}},\ }\bibfield  {title} {\bibinfo
  {title} {Collisional quantum thermometry},\ }\href
  {https://doi.org/10.1103/PhysRevLett.123.180602} {\bibfield  {journal}
  {\bibinfo  {journal} {Phys. Rev. Lett.}\ }\textbf {\bibinfo {volume} {123}},\
  \bibinfo {pages} {180602} (\bibinfo {year} {2019})}\BibitemShut {NoStop}%
\bibitem [{\citenamefont {Paris}(2009{\natexlab{a}})}]{Colisao_Paris}%
  \BibitemOpen
  \bibfield  {author} {\bibinfo {author} {\bibfnamefont {M.~G.~A.}\
  \bibnamefont {Paris}},\ }\bibfield  {title} {\bibinfo {title} {Quantum
  estimation for quantum technology},\ }\href
  {https://doi.org/10.1142/S0219749909004839} {\bibfield  {journal} {\bibinfo
  {journal} {Int. J. Quantum. Inform.}\ }\textbf {\bibinfo {volume} {7}},\
  \bibinfo {pages} {125} (\bibinfo {year} {2009}{\natexlab{a}})}\BibitemShut
  {NoStop}%
\bibitem [{\citenamefont {Shu}\ \emph {et~al.}(2020)\citenamefont {Shu},
  \citenamefont {Seah},\ and\ \citenamefont {Scarani}}]{Colisao_Scarani}%
  \BibitemOpen
  \bibfield  {author} {\bibinfo {author} {\bibfnamefont {A.}~\bibnamefont
  {Shu}}, \bibinfo {author} {\bibfnamefont {S.}~\bibnamefont {Seah}},\ and\
  \bibinfo {author} {\bibfnamefont {V.}~\bibnamefont {Scarani}},\ }\bibfield
  {title} {\bibinfo {title} {Surpassing the thermal {C}ram\'er-{R}ao bound with
  collisional thermometry},\ }\href
  {https://doi.org/10.1103/PhysRevA.102.042417} {\bibfield  {journal} {\bibinfo
   {journal} {Phys. Rev. A}\ }\textbf {\bibinfo {volume} {102}},\ \bibinfo
  {pages} {042417} (\bibinfo {year} {2020})}\BibitemShut {NoStop}%
\bibitem [{\citenamefont {Breuer}\ \emph {et~al.}(2009)\citenamefont {Breuer},
  \citenamefont {Laine},\ and\ \citenamefont {Piilo}}]{Breuer2009}%
  \BibitemOpen
  \bibfield  {author} {\bibinfo {author} {\bibfnamefont {H.-P.}\ \bibnamefont
  {Breuer}}, \bibinfo {author} {\bibfnamefont {E.-M.}\ \bibnamefont {Laine}},\
  and\ \bibinfo {author} {\bibfnamefont {J.}~\bibnamefont {Piilo}},\ }\bibfield
   {title} {\bibinfo {title} {Measure for the degree of non-markovian behavior
  of quantum processes in open systems},\ }\href
  {https://doi.org/10.1103/PhysRevLett.103.210401} {\bibfield  {journal}
  {\bibinfo  {journal} {Phys. Rev. Lett.}\ }\textbf {\bibinfo {volume} {103}},\
  \bibinfo {pages} {210401} (\bibinfo {year} {2009})}\BibitemShut {NoStop}%
\bibitem [{\citenamefont {Laine}\ \emph {et~al.}(2010)\citenamefont {Laine},
  \citenamefont {Piilo},\ and\ \citenamefont {Breuer}}]{Breuer2010}%
  \BibitemOpen
  \bibfield  {author} {\bibinfo {author} {\bibfnamefont {E.-M.}\ \bibnamefont
  {Laine}}, \bibinfo {author} {\bibfnamefont {J.}~\bibnamefont {Piilo}},\ and\
  \bibinfo {author} {\bibfnamefont {H.-P.}\ \bibnamefont {Breuer}},\ }\bibfield
   {title} {\bibinfo {title} {Measure for the non-markovianity of quantum
  processes},\ }\href {https://doi.org/10.1103/PhysRevA.81.062115} {\bibfield
  {journal} {\bibinfo  {journal} {Phys. Rev. A}\ }\textbf {\bibinfo {volume}
  {81}},\ \bibinfo {pages} {062115} (\bibinfo {year} {2010})}\BibitemShut
  {NoStop}%
\bibitem [{\citenamefont {Mendon\c{c}a}\ \emph {et~al.}(2024)\citenamefont
  {Mendon\c{c}a}, \citenamefont {C\'eleri}, \citenamefont {Paternostro},\ and\
  \citenamefont {Soares-Pinto}}]{mendonca2024InfFlow}%
  \BibitemOpen
  \bibfield  {author} {\bibinfo {author} {\bibfnamefont {T.~M.}\ \bibnamefont
  {Mendon\c{c}a}}, \bibinfo {author} {\bibfnamefont {L.~C.}\ \bibnamefont
  {C\'eleri}}, \bibinfo {author} {\bibfnamefont {M.}~\bibnamefont
  {Paternostro}},\ and\ \bibinfo {author} {\bibfnamefont {D.~O.}\ \bibnamefont
  {Soares-Pinto}},\ }\bibfield  {title} {\bibinfo {title} {System-environment
  quantum information flow},\ }\href
  {https://doi.org/10.1103/PhysRevA.110.L040401} {\bibfield  {journal}
  {\bibinfo  {journal} {Phys. Rev. A}\ }\textbf {\bibinfo {volume} {110}},\
  \bibinfo {pages} {L040401} (\bibinfo {year} {2024})}\BibitemShut {NoStop}%
\bibitem [{\citenamefont {Breuer}\ and\ \citenamefont
  {Petruccione}(2002)}]{PetruccioneLivro2002}%
  \BibitemOpen
  \bibfield  {author} {\bibinfo {author} {\bibfnamefont {H.~P.}\ \bibnamefont
  {Breuer}}\ and\ \bibinfo {author} {\bibfnamefont {F.}~\bibnamefont
  {Petruccione}},\ }\href@noop {} {\emph {\bibinfo {title} {The theory of open
  quantum systems}}}\ (\bibinfo  {publisher} {Oxford University Press},\
  \bibinfo {address} {New York},\ \bibinfo {year} {2002})\BibitemShut {NoStop}%
\bibitem [{\citenamefont {Paris}(2009{\natexlab{b}})}]{Paris2009}%
  \BibitemOpen
  \bibfield  {author} {\bibinfo {author} {\bibfnamefont {M.~G.~A.}\
  \bibnamefont {Paris}},\ }\bibfield  {title} {\bibinfo {title} {Quantum
  estimation for quantum technology},\ }\href@noop {} {\bibfield  {journal}
  {\bibinfo  {journal} {Int. J. Quantum Inf}\ }\textbf {\bibinfo {volume} {7}}
  (\bibinfo {year} {2009}{\natexlab{b}})}\BibitemShut {NoStop}%
\bibitem [{\citenamefont {Breuer}\ \emph {et~al.}(2016)\citenamefont {Breuer},
  \citenamefont {Laine}, \citenamefont {Piilo},\ and\ \citenamefont
  {Vacchini}}]{Breuer2016_colloquium}%
  \BibitemOpen
  \bibfield  {author} {\bibinfo {author} {\bibfnamefont {H.-P.}\ \bibnamefont
  {Breuer}}, \bibinfo {author} {\bibfnamefont {E.-M.}\ \bibnamefont {Laine}},
  \bibinfo {author} {\bibfnamefont {J.}~\bibnamefont {Piilo}},\ and\ \bibinfo
  {author} {\bibfnamefont {B.}~\bibnamefont {Vacchini}},\ }\bibfield  {title}
  {\bibinfo {title} {Colloquium: Non-markovian dynamics in open quantum
  systems},\ }\href {https://doi.org/10.1103/RevModPhys.88.021002} {\bibfield
  {journal} {\bibinfo  {journal} {Rev. Mod. Phys.}\ }\textbf {\bibinfo {volume}
  {88}},\ \bibinfo {pages} {021002} (\bibinfo {year} {2016})}\BibitemShut
  {NoStop}%
\bibitem [{\citenamefont {Rivas}\ \emph {et~al.}(2014)\citenamefont {Rivas},
  \citenamefont {Huelga},\ and\ \citenamefont {Plenio}}]{Rivas2014}%
  \BibitemOpen
  \bibfield  {author} {\bibinfo {author} {\bibfnamefont {A.}~\bibnamefont
  {Rivas}}, \bibinfo {author} {\bibfnamefont {S.~F.}\ \bibnamefont {Huelga}},\
  and\ \bibinfo {author} {\bibfnamefont {M.~B.}\ \bibnamefont {Plenio}},\
  }\bibfield  {title} {\bibinfo {title} {Quantum non-markovianity:
  characterization, quantification and detection},\ }\href
  {https://doi.org/10.1088/0034-4885/77/9/094001} {\bibfield  {journal}
  {\bibinfo  {journal} {Reports on Progress in Physics}\ }\textbf {\bibinfo
  {volume} {77}},\ \bibinfo {pages} {094001} (\bibinfo {year}
  {2014})}\BibitemShut {NoStop}%
\bibitem [{\citenamefont {Li}\ \emph {et~al.}(2018)\citenamefont {Li},
  \citenamefont {Hall},\ and\ \citenamefont {Wiseman}}]{Li2018}%
  \BibitemOpen
  \bibfield  {author} {\bibinfo {author} {\bibfnamefont {L.}~\bibnamefont
  {Li}}, \bibinfo {author} {\bibfnamefont {M.~J.}\ \bibnamefont {Hall}},\ and\
  \bibinfo {author} {\bibfnamefont {H.~M.}\ \bibnamefont {Wiseman}},\
  }\bibfield  {title} {\bibinfo {title} {Concepts of quantum non-markovianity:
  A hierarchy},\ }\href
  {https://doi.org/https://doi.org/10.1016/j.physrep.2018.07.001} {\bibfield
  {journal} {\bibinfo  {journal} {Physics Reports}\ }\textbf {\bibinfo {volume}
  {759}},\ \bibinfo {pages} {1} (\bibinfo {year} {2018})},\ \bibinfo {note}
  {concepts of quantum non-Markovianity: A hierarchy}\BibitemShut {NoStop}%
\bibitem [{\citenamefont {Chr\'uci\'nski}(2022)}]{Dariusz2022}%
  \BibitemOpen
  \bibfield  {author} {\bibinfo {author} {\bibfnamefont {D.}~\bibnamefont
  {Chr\'uci\'nski}},\ }\bibfield  {title} {\bibinfo {title} {{D}ynamical maps
  beyond markovian regime},\ }\href
  {https://doi.org/https://doi.org/10.1016/j.physrep.2022.09.003} {\bibfield
  {journal} {\bibinfo  {journal} {Physics Reports}\ }\textbf {\bibinfo {volume}
  {992}},\ \bibinfo {pages} {1} (\bibinfo {year} {2022})}\BibitemShut {NoStop}%
\bibitem [{\citenamefont {Nielsen}\ and\ \citenamefont
  {Chuang}(2010)}]{NielsenLivro2010}%
  \BibitemOpen
  \bibfield  {author} {\bibinfo {author} {\bibfnamefont {M.~A.}\ \bibnamefont
  {Nielsen}}\ and\ \bibinfo {author} {\bibfnamefont {I.~L.}\ \bibnamefont
  {Chuang}},\ }\href@noop {} {\emph {\bibinfo {title} {Quantum Computation and
  Quantum Information}}},\ \bibinfo {edition} {10th}\ ed.\ (\bibinfo
  {publisher} {Cambridge University Press},\ \bibinfo {address} {New York},\
  \bibinfo {year} {2010})\BibitemShut {NoStop}%
\end{thebibliography}%

\end{document}